\documentclass[pra,aps,groupedaddress,twocolumn,floatfix,nofootinbib]{revtex4}
\usepackage{amsmath,amsthm,amsfonts,amssymb,graphics,graphicx,epsfig,color,times,natbib,subfigure}
\usepackage{bbm} 

\newcommand{\ket}[1]{\mbox{$ | #1 \rangle $}}

\newcommand{\be}{\begin{equation}}
\newcommand{\ee}{\end{equation}}
\newcommand{\ba}{\begin{eqnarray}}
\newcommand{\ea}{\end{eqnarray}}

\newcommand{\demi}{\frac{1}{2}}

\newcommand{\real}{\begin{picture}(8,8)\put(0,0){R}\put(0,0){\line(0,1){7}}\end{picture}}

\newcommand{\one}{\leavevmode\hbox{\small1\normalsize\kern-.33em1}}

\begin{document}

\title{Detection Loophole in Bell experiments: \\ How post-selected local correlations can look non-local}
\author{Cyril Branciard}
\affiliation{Centre for Quantum Computer Technology, School of Mathematics and Physics, The University of Queensland, St Lucia, QLD 4072, Australia}
\date{\today}

\begin{abstract}
A common problem in Bell type experiments is the well-known detection loophole: if the detection efficiencies are not perfect and if one simply post-selects the conclusive events, one might observe a violation of a Bell inequality, even though a local model could have explained the experimental results. In this paper, we analyze the set of all post-selected correlations that can be explained by a local model, and show that it forms a polytope, 
larger than the Bell local polytope. We characterize the facets of this \emph{post-selected local polytope} in the CHSH scenario, where two parties have binary inputs and outcomes. Our approach gives new insights on the detection loophole problem.

\end{abstract}

\maketitle

%

\section{Introduction}


Quantum non-locality, i.e. the fact that, in Bell's terminology, no locally causal explanation can be given to quantum mechanical correlations~\cite{bell}, is certainly one of the most fascinating and intriguing features of the quantum theory. Our classical understanding and apprehension of the physical world is quite disrupted by this characteristic, and experimental demonstrations are necessary for physicists and philosophers to accept such an upheaval.

A signature of non-locality is the violation of a Bell inequality~\cite{bell}. In the last 30 years, many Bell-type experiments have been performed to demonstrate quantum non-locality~\cite{bell_experiments}, all of them showing good agreement with the quantum predictions. However, none of these experiments can be considered as perfectly convincing, as so far they all suffer from persistent loopholes: the sceptic can always find a (more or less far-fetched) classical explanation for the observed data.
Given the crucial role of non-locality in quantum information processing applications~\cite{cleve_buhrman,DIQKD,bardyn,pironio_RNG}, loophole-free demonstrations of quantum non-locality are highly desirable.

One of these loopholes is known as the detection loophole~\cite{pearle_detLH}. Typically, in photonic experiments the detection efficiencies are not perfect, and one usually post-selects the detected events to show a violation of a Bell inequality. However, there might exist a model that exploits the detector inefficiencies to reproduce the experimental data~\cite{pearle_detLH,gisin_gisin}, in perfect agreement with Bell's assumption of local causality~\cite{bell}. In order to circumvent this problem, one usually resorts to the fair sampling assumption, that the detected particles are representative of all those emitted from the source, but this additional assumption is certainly not satisfactory. 
Closing the detection loophole would require either improving the detection efficiencies of the detectors used in Bell experiments, or finding Bell inequalities that are more robust to detection inefficiencies, as reported in~\cite{massar2002,massar_pironio_roland_gisin,massar_pironio,brunner_gisin,pal_vertesi,vertesi_pironio_brunner}. Although the known necessary detection efficiencies are still quite high, a photonic detection-loophole-free Bell experiment seems possible in the near future.





Our goal here is to improve our understanding of the detection loophole problem and get a better intuition on it, by studying how post-selection modifies the requirements for demonstrating non-locality. We will show that the set of post-selected local correlations is a polytope, that includes the Bell local polytope (Sec~\ref{sec_Lps}). To illustrate this, we will consider the CHSH scenario (Sec~\ref{sec_Lps_CHSH}), with two parties both having two possible inputs and two outcomes (excluding the no-detection outcomes).
This approach gives new insights on the (non-)~locality of post-selected correlations. It will allow us in particular to re-derive and prove the optimality of Eberhard's result on the tolerance of Bell tests to detection inefficiencies~\cite{eberhard} in the CHSH scenario, and to understand why the quantum correlation that gives the largest violation of the CHSH inequality~\cite{chsh} is not the most robust to detection inefficiencies.


\section{Post-selected local correlations}

\label{sec_Lps}

\subsection{Bell-type experiment with imperfect detection efficiencies}

Let us consider a typical Bell type experiment involving two parties, Alice and Bob, with $m_A$ and $m_B$ inputs, and $n_A$ and $n_B$ outcomes respectively~\footnote{In full generality, we could consider different numbers of outcomes for each observable, and a larger number of parties as well. The following study can easily be adapted to these cases.}.

If Alice and Bob have non perfect detection efficiencies, we need to also take into account the possibility for Alice and Bob's detectors not to fire (``$\emptyset$"). They will thus actually have, respectively, $n_A+1$ and $n_B+1$ possible outcomes, denoted
\ba
a = 1, \dots, n_A, \emptyset \ ; \quad
b = 1, \dots, n_B, \emptyset \, .
\ea

After repeating the experiment many times, Alice and Bob can estimate their correlations, i.e. the 
probability distribution
\ba
P_0(a,b|x,y)
\ea
for $a = 1, \dots, n_A, \emptyset$,  $b = 1, \dots, n_B, \emptyset$, and for the choice of measurement settings $x = 1, \dots, m_A$ and $y = 1, \dots, m_B$.
We call $P_0$ the {\it a priori} correlation: it is estimated before post-selection. As it is standard in the study of non-locality, we will assume that $P_0$ is non-signaling (i.e., $P_0(a|x,y) = P_0(a|x)$ and $P_0(b|x,y) = P_0(b|y)$); in an experiment, this can in particular be ensured by having Alice and Bob space-like separated.

\bigskip

We will assume in the following that Alice's and Bob's detection probabilities are independent of their choice of measurement setting, and of what happens on the other partner's side. Defining $\eta_A$ (resp. $\eta_B$) to be Alice's (resp. Bob's) detection efficiency, this translates into the following constraints:
\ba
\forall \ b,x,y, && P_0(a\neq\emptyset,b|x,y) = \eta_A P_0(b|y) \label{eq_etaA} \\
\forall \ a,x,y, && P_0(a,b\neq\emptyset|x,y) = \eta_B P_0(a|x) \label{eq_etaB}\, ,
\ea
where we write $P_0(a\neq\emptyset,b|x,y) = \sum_{a\neq\emptyset}P_0(a,b|x,y)$, $P_0(a,b\neq\emptyset|x,y) = \sum_{b\neq\emptyset}P_0(a,b|x,y)$, and where we used the no-signaling assumption.

This implies in particular, that (with obvious notations)
\ba
\forall \ x,y, && P_0(a\neq\emptyset,b\neq\emptyset|x,y) = \eta_A \eta_B \, .
\ea

\subsection{Post-selected correlations}

From their experimental data, Alice and Bob can post-select the conclusive events, when both detected their particle, and discard the non-conclusive events, as soon as one of the particles was not detected. They can thus estimate their post-selected correlations, now for $a = 1, \dots, n_A$ and $b = 1, \dots, n_B$:
\ba
P_{ps}(a,b|x,y) &=& P_0(a,b|x,y,a\neq\emptyset,b\neq\emptyset) \nonumber \\
&=& \frac{P_0(a,b|x,y)}{P_0(a\neq\emptyset,b\neq\emptyset|x,y)} 
\ea
i.e.
\ba
P_{ps}(a,b|x,y) = \frac{1}{\eta_A \eta_B}P_0(a,b|x,y) \, .
\label{eq_Pps}
\ea
Note that the preceding independence assumption for $\eta_A$ and $\eta_B$ ensures that $P_{ps}$ is also non-signaling.

\subsection{Local causality assumption}

In order for Alice and Bob to demonstrate non-locality in their experiment, they need to check if their data {\it before post-selection} can be explained by a local model.

The {\it a priori} correlation $P_0(a,b|x,y)$ satisfies Bell's standard local causality assumption~\cite{bell} if it can be decomposed in the form \ba
P_0(a,b|x,y) = \int d\lambda \, \rho(\lambda) \ P_0(a|x,\lambda) P_0(b|y,\lambda) \, , \label{def_loc}
\ea
for some local variables $\lambda$ distributed according to $\rho(\lambda)$. It is well known that the set of local correlations forms a convex 
polytope~\cite{pitowsky} (which we call, in our case here, the ``local {\it a priori} polytope'', and denote by ${\cal L}_0$), included in the polytope that contains all non-signaling correlations (the ``non-signaling {\it a priori} polytope'', denoted by ${\cal P}_0$).

\bigskip

Coming back to the post-selected correlation $P_{ps}$, we'll say that it is ``post-selected local'' 
 if it can be obtained by post-selecting the conclusive events of a local {\it a priori} correlation $P_0$ satisfying (\ref{def_loc}).

From eqs (\ref{eq_etaA}-\ref{eq_etaB}) and (\ref{eq_Pps}), one can see that the set of post-selected local correlations is, up to a factor $\frac{1}{\eta_A\eta_B}$, the intersection of the local {\it a priori} polytope ${\cal L}_0$, with the subspace defined by eqs (\ref{eq_etaA}-\ref{eq_etaB}). The intersection of a polytope with a subspace being a polytope~\cite{ziegler}, the set of post-selected local correlations is thus also a polytope, which we denote by ${\cal L}_{ps}(\eta_A,\eta_B)$ (or simply ${\cal L}_{ps}$ for short).

\bigskip

The post-selected local polytope ${\cal L}_{ps}$ clearly includes the local polytope ${\cal L}$, that contains the local probability distributions for $m_A$ and $m_B$ inputs, and $n_A$ and $n_B$ outcomes~\footnote{Any local correlation $P \in {\cal L}$ can indeed be turned into a local {\it a priori} correlation $P_0 \in {\cal L}_0$ by just adding the possibility for Alice and Bob's detectors not to fire, with independent probabilities $\eta_A$ and $\eta_B$. After post-selection from $P_0$, we obtain back $P_{ps} = P$, which proves that $P \in {\cal L}_{ps}$.
\\
A similar argument allows one to show, more generally, that ${\cal L}_{ps}(\eta_A,\eta_B) \subset {\cal L}_{ps}(\eta_A',\eta_B')$ for any $\eta_A' \leq \eta_A$ and $\eta_B' \leq \eta_B$. Note that ${\cal L} = {\cal L}_{ps}(\eta_A=1,\eta_B=1)$.}; both are included in the corresponding non-signaling polytope ${\cal P}$~\footnote{The local and non-signaling polytopes ${\cal L}$ and ${\cal P}$ should not be confused with the previous local and non-signaling ``{\it a priori} polytopes'' ${\cal L}_0$ and ${\cal P}_0$: the latter were indeed defined for correlations with $m_A$ and $m_B$ inputs, and $n_A+1$ and $n_B+1$ outcomes.
\\
In general, ${\cal L}_0$ and ${\cal P}_0$ are of dimension $(m_An_A+1)(m_Bn_B+1)-1$, while ${\cal L}$, ${\cal L}_{ps}$ and ${\cal P}$ are of dimension $(m_A(n_A-1)+1)(m_B(n_B-1)+1)-1$~\cite{collins_gisin}.}. However, there can be correlations in ${\cal L}_{ps}$ that are not in ${\cal L}$: these correlations will violate the standard Bell inequalities (which delimit the polytope ${\cal L}$) and therefore might ``look non-local'', but they can still be explained by a local model with post-selection~\footnote{In fact, to conclude that these correlations are indeed non-local, one would usually resort to the fair sampling assumption; we don't want to use this additional assumption here.}.

Studying and characterizing the polytope ${\cal L}_{ps}(\eta_A,\eta_B)$ allows one to understand which post-selected correlations can or cannot be explained by a local model. This can easily be done once the polytope ${\cal L}_0$ has been characterized: indeed, the facets of ${\cal L}_0$ define, of course, valid inequalities for the intersection of ${\cal L}_0$ with the subspace defined by eqs (\ref{eq_etaA}-\ref{eq_etaB}); using (\ref{eq_etaA}-\ref{eq_etaB}) and (\ref{eq_Pps}), this leads to valid inequalities for the post-selected local probabilities $P_{ps} \in {\cal L}_{ps}(\eta_A,\eta_B)$.
These inequalities are not all facets of ${\cal L}_{ps}$, but since the polytope ${\cal L}_{ps}$ is precisely delimited by the facets of ${\cal L}_0$, all of its own facets must be in the list of valid inequalities just obtained. Sorting all these inequalities thus allows one to extract all the facets of ${\cal L}_{ps}$.

In the following, we illustrate this in the CHSH scenario, where Alice and Bob have two possible inputs with binary outcomes (plus the no-detection events).



\section{Post-selected local polytope ${\cal L}_{ps}(\eta_A,\eta_B)$ \newline in the CHSH scenario}

\label{sec_Lps_CHSH}

\subsection{The standard CHSH scenario: \\ 2 inputs, 2 outcomes for Alice and Bob}

The CHSH scenario corresponds to the simplest case, where Alice and Bob can both choose between two measurement settings, and have binary outcomes. In this case, all the non-trivial Bell inequalities that delimit the local polytope $\cal L$ are equivalent to the CHSH inequality\footnote{Two inequalities are equivalent if they can be transformed into one another by relabeling the inputs, the outcomes, and/or exchanging the parties. In our case here, there are 8 different equivalent versions of CHSH. The local polytope ${\cal L}$ also has 16 other (equivalent) facets, which simply correspond to the non-negativity of the probabilities $P(a,b|x,y)$; these facets, and the corresponding inequalities, are said to be trivial.}~\cite{fine,chsh}, which can be written in the CH form~\cite{ch} as:
\ba
P(11|11)+P(11|12)+P(11|21) \quad \nonumber \\
-P(11|22)-P_A(1|1)-P_B(1|1) & \leq & 0
\ea
where $P_A(a|x)$ (resp. $P_B(b|y)$) denotes the marginal probability distribution of Alice (resp. Bob).

It is convenient to use the notation introduced in~\cite{collins_gisin}, and write the CHSH (or CH) inequality as
\ba
 I_{CH} \, =\ 
 \begin{array}{c||cc} 
 & -1 & 0 \\ \hline \hline
 -1 & 1 & 1 \\
 0 & 1 & -1
 \end{array}
 \ \leq \ 0
 \label{CHSH}
\ea
where the coefficients in the table are those that appear in front of the probabilities of getting the first outcome:
\ba
 \begin{array}{c||cc} 
 & P_B(1|1) & P_B(1|2) \\ \hline \hline
 P_A(1|1) & P(11|11) & P(11|12) \\
 P_A(1|2) & P(11|21) & P(11|22)
 \end{array} \ .
 \label{def_notation}
\ea

\subsection{The CHSH scenario with inefficient detectors: \\ 2 inputs, 3 outcomes for Alice and Bob}

In the case of inefficient detectors, there are now 3 possible outcomes on Alice and Bob's sides: $\emptyset, 1$ and 2.

The polytope ${\cal L}_0$, corresponding to 2 inputs and 3 outcomes for both Alice and Bob, has been fully characterized in~\cite{masanes,collins_gisin}. It has 1116 facets, of which 36 are trivial (simply corresponding to non-negative probabilities), 648 are of the CHSH form~\footnote{In fact, there are two inequivalent sets of 324 equivalent CHSH-like inequalities each.} 
(with 2 outcomes grouped together on each side, so that Alice and Bob both have only two effective outcomes), and 432 are equivalent to the CGLMP inequality~\cite{CGLMP}:
\ba
 I_{CGLMP} \, =\ 
 \begin{array}{c||cc|cc} 
 & -1 & -1 & 0 & 0 \\ \hline \hline
 -1 & 1 & 0 & 1 & 1 \\
 -1 & 1 & 1 & 0 & 1 \\ \hline
 0 & 1 & 1 & -1 & -1 \\
 0 & 0 & 1 & 0 & -1
 \end{array}
 \ \leq \ 0 \label{def_CGLMP}
 \ea
(where the rows now correspond to $a_{x=1}=1$, $a_{x=1}=2$, $a_{x=2}=1$, $a_{x=2}=2$, and similarly for the columns~\cite{collins_gisin}).

\subsection{Bell inequalities for ${\cal L}_{ps}(\eta_A,\eta_B)$}


From the 1116 facets of ${\cal L}_0$, and using (\ref{eq_etaA}-\ref{eq_etaB}) and (\ref{eq_Pps}), we obtain a list of valid inequalities for ${\cal L}_{ps}(\eta_A,\eta_B)$ in the CHSH scenario. After sorting them, we find that, in addition to the trivial inequalities, it is actually sufficient to consider only the 64 equivalent forms of the following ones, as all the other inequalities are either trivial, or can be derived from them (see Appendix~\ref{app_all_ineqs}):
\ba
& -1 \ \leq \ I_{CH}^{\eta_A,\eta_B}
 \ \leq \ 0 \label{CH_ps} \\
& {\mathrm{with}}
 \quad
 I_{CH}^{\eta_A,\eta_B} \ = \
 \begin{array}{c||cc} 
 & -\eta_B & 0 \\ \hline \hline
 -\eta_A & \eta_A \eta_B & \eta_A \eta_B \\
 0 & \eta_A \eta_B & -\eta_A \eta_B
 \end{array}
 \, .
\ea

Interestingly, the above inequalities are simply obtained from the CH inequality, by grouping, for each observable, the outcome ``$\emptyset$" with one of the other outcomes~\footnote{Note that the two inequalities in (\ref{CH_ps}) are in general non-equivalent, except if $\eta_A = 1$ or $\eta_B = 1$.} (see~\cite{eberhard,cabello_asym,brunner_asym} for previous derivations of the inequality $I_{CH}^{\eta_A,\eta_B} \leq 0$). Here, the CGLMP inequality does not provide any additional Bell inequalities for the CHSH scenario with imperfect detectors.

\bigskip

We prove in Appendix~\ref{app_facets} that all the facets of ${\cal L}_{ps}(\eta_A,\eta_B)$ are, precisely, either of the trivial form $P_{ps}(a,b|x,y) \geq 0$, or of the form $I_{CH}^{\eta_A,\eta_B} \leq 0$ (if $\eta_A+\eta_B < 3\eta_A\eta_B$), or $I_{CH}^{\eta_A,\eta_B} \geq -1$ (under a stronger constraint $h(\eta_A,\eta_B) < 0$, with $h$ defined in~(\ref{eq_def_fgh})).


With this characterization, we now have the full list of all facets of ${\cal L}_{ps}$; one can then easily check if a given correlation is ``post-selected local'' or not. For that, Bell inequalities of the form (\ref{CH_ps}) should be tested rather than the standard CHSH inequality (\ref{CHSH}).


\subsection{Application: necessary conditions on $\eta_A,\eta_B$ \\ to observe non-locality}

One can now easily derive necessary conditions on $\eta_A,\eta_B$ to observe non-locality. Indeed, as proven in Appendix~\ref{app_facets}, in order for ${\cal L}_{ps}$ to have non-trivial facets, one must have
\be \eta_A+\eta_B < 3\eta_A\eta_B \, . \label{constr_etas} \ee
If this constraint is not satisfied, then only trivial inequalities delimit ${\cal L}_{ps}$
(which is then actually equal to the full non-signaling polytope ${\cal P}$), and no violation can be observed.

In the symmetric case $\eta_A=\eta_B=\eta$, we get the necessary condition
\be \eta > \frac{2}{3} \, , \ee
which corresponds to the threshold obtained by Eberhard~\cite{eberhard}. The condition (\ref{constr_etas}), for general values of $\eta_A$ and $\eta_B$, had also been derived previously in~\cite{cabello_asym}. For the special case $\eta_A=1$, we get the constraint $\eta_B > \demi$ (see also~\cite{brunner_asym}).

\bigskip

All these previous derivations~\cite{eberhard,cabello_asym,brunner_asym} were based on the inequality $I_{CH}^{\eta_A,\eta_B} \leq 0$. Our approach here allows us to justify this choice: we prove that this is, together with $I_{CH}^{\eta_A,\eta_B} \geq -1$, the only relevant inequality in a CHSH scenario with imperfect detection efficiencies, but that $I_{CH}^{\eta_A,\eta_B} \geq -1$ is less robust to detection inefficiencies~\footnote{As proven in Appendix~\ref{app_facets}, the inequality $I_{CH}^{\eta_A,\eta_B} \geq -1$ is a facet of ${\cal L}_{ps}$ only if $h(\eta_A,\eta_B) < 0$, which is more restrictive than (\ref{constr_etas}).}.

In comparison to previous proofs, we really derived here a necessary condition for observing non-locality in a CHSH scenario, not only for observing a violation of a given inequality. To our knowledge, only the conditions $\eta>\frac{2}{3}$ in the symmetric case and $\eta_B>\demi$ in the special asymmetric case (with $\eta_A=1$) were known to be necessary to observe non-locality~\cite{massar_pironio}.

\bigskip

Let us finally mention that this necessary condition is valid for all non-signaling theories, and is not limited to quantum mechanics. Whether it is also a sufficient condition does however depend on the correlations one can achieve. It turns out that this is indeed the case for quantum correlations, which can violate $I_{CH}^{\eta_A,\eta_B} \leq 0$ for all $\eta_A,\eta_B$ such that $\eta_A+\eta_B < 3\eta_A\eta_B$~\cite{cabello_asym}.

\subsection{Geometric views}

In order to get a better intuition, we now illustrate what the post-selected local polytope ${\cal L}_{ps}(\eta_A,\eta_B)$ looks like in some particular two-dimensional slices of the correlation space.

\subsubsection{A nicely symmetric 2-D slice}

Let us first consider the 2-D slice that contains two (equivalent) PR boxes~\cite{PR} $P_{PR}$ and $P_{PR'}$, and the fully random correlation $P_r$, defined as follows, in the notation of (\ref{def_notation}):
\ba
&
 P_{PR} =
 \begin{array}{c||cc} 
 & 1/2 & 1/2 \\ \hline \hline
 1/2 & 1/2 & 1/2 \\
 1/2 & 1/2 & 0
 \end{array} \ , \quad
 P_{PR'} = 
 \begin{array}{c||cc} 
 & 1/2 & 1/2 \\ \hline \hline
 1/2 & 0 & 1/2 \\
 1/2 & 1/2 & 1/2
 \end{array} \ , \nonumber \\
 &
 P_r =
 \begin{array}{c||cc} 
 & 1/2 & 1/2 \\ \hline \hline
 1/2 & 1/4 & 1/4 \\
 1/2 & 1/4 & 1/4
 \end{array} \, .
\ea
Any correlation in this slice can then be written in the form
\ba
P_{xy} = x P_{PR'} + y P_{PR} + (1-x-y) P_r \, ,
\ea
with $x,y \in \real$.

\bigskip

The trivial facets, the inequalities $I_{CH}^{\eta_A,\eta_B} \leq 0$ and $I_{CH}^{\eta_A,\eta_B} \geq -1$ (together with all their equivalent versions) respectively impose the following constraints on $x$ and $y$ for $P_{xy}$ to be in ${\cal L}_{ps}(\eta_A,\eta_B)$:
\ba
|x+y| & \leq & 1 \\
|x|,|y| & \leq & \frac{\eta_A+\eta_B-\eta_A\eta_B}{2\eta_A\eta_B} := F(\eta_A,\eta_B) \\
|x|,|y| & \leq & \frac{2-\eta_A-\eta_B+\eta_A\eta_B}{2\eta_A\eta_B} := G(\eta_A,\eta_B) \, .
\ea
Note that $\demi \leq F(\eta_A,\eta_B) \leq G(\eta_A,\eta_B)$,
and therefore the last inequality above is implied by the previous one.

\begin{figure}
\begin{center}
\epsfxsize=8cm
\epsfbox{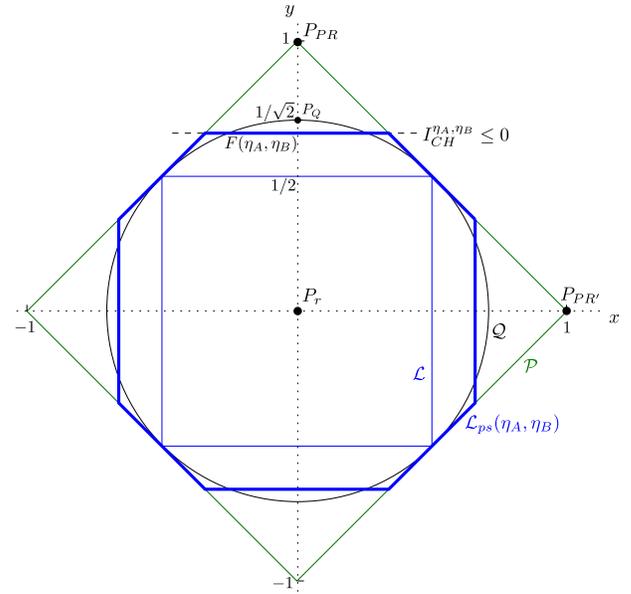}
\caption{(Color online.) Two-dimensional slice of the correlation space corresponding to the CHSH scenario, containing the correlations $P_{PR}$, $P_{PR'}$ and $P_r$. ${\cal L}_{ps}(\eta_A,\eta_B)$ is the thick blue polytope; the inner blue square delimits the local polytope ${\cal L} = {\cal L}_{ps}(1,1)$; the outer green diamond delimits the no-signaling polytope ${\cal P}$; the black circle corresponds to the set ${\cal Q}$ of quantum correlations.}
\label{fig_slice_PR_PR2}
\end{center}
\end{figure}

\bigskip

The structure of this two-dimensional slice, with these delimiting inequalities, is illustrated on Figure~\ref{fig_slice_PR_PR2}.

The set ${\cal Q}$ of quantum correlations corresponds in this slice to the disk $x^2+y^2 \leq \demi$~\footnote{The constraint $x^2+y^2 \leq \demi$ can be obtained from the criteria derived in~\cite{masanes_Qset} (see also~\cite{navascues}). The bound is tight, which can be seen as follows: consider the standard CHSH settings $\vec a_1 = \vec z$, $\vec a_2 = \vec x$ and $\vec b_1, \vec b_2 = \frac{\vec z \pm \vec x}{\sqrt{2}}$ (represented as vectors on the Bloch sphere), measured on the maximally entangled state $\ket{\Phi^+}$; we obtain the correlation $P_Q$, corresponding to $x=0,y=\frac{1}{\sqrt{2}}$. Now, rotate the two settings of Bob together in the $xz$ plane of the Bloch sphere, and the whole circle $x^2+y^2=\demi$ is recovered.}. We can thus see that \emph{in this slice}, a violation of the inequality $I_{CH}^{\eta_A,\eta_B} \leq 0$ can be obtained quantum mechanically only if $F(\eta_A,\eta_B) < 1/\sqrt{2}$, and is maximal for the correlation $P_Q$ that maximizes the violation of the standard CHSH inequality (i.e., with the standard choice of the CHSH settings, measured on a maximally entangled state). In the symmetric case $\eta_A=\eta_B=\eta$, we get $\eta > 2\sqrt{2}-2\simeq 83\%$, which corresponds to the bound derived in~\cite{garg_mermin}.

\subsubsection{Illustration of Eberhard's result}

Eberhard's result~\cite{eberhard}, that the correlation that gives a maximal violation of the standard CHSH inequality is not the most robust to detection inefficiencies, may seem surprising. Our approach here allows us to get a geometric intuition and a better understanding of this result. 

As the detection efficiencies $\eta_A$ and/or $\eta_B$ decrease, the polytope ${\cal L}_{ps}(\eta_A,\eta_B)$ continuously gets bigger, until it becomes equal to the full non-signaling polytope $\cal P$ when $\eta_A+\eta_B \geq 3\eta_A\eta_B$. Just before reaching the size of $\cal P$, i.e., for $3\eta_A\eta_B-\eta_A-\eta_B$ just slightly positive, the last correlations that are non-``post-selected local" are therefore to be found close to the boundaries of $\cal P$; and as already mentioned, whatever $\eta_A,\eta_B$ such that $\eta_A+\eta_B < 3\eta_A\eta_B$, there exists quantum correlations in ${\cal P} \setminus {\cal L}_{ps}$~\cite{cabello_asym}.
Clearly, the quantum correlation $P_Q$ is not close to the boundary of $\cal P$ (see Figure~\ref{fig_slice_PR_PR2}), and we now understand why it is not the most robust to detection efficiency.

\begin{figure}
\begin{center}
\epsfxsize=8cm
\epsfbox{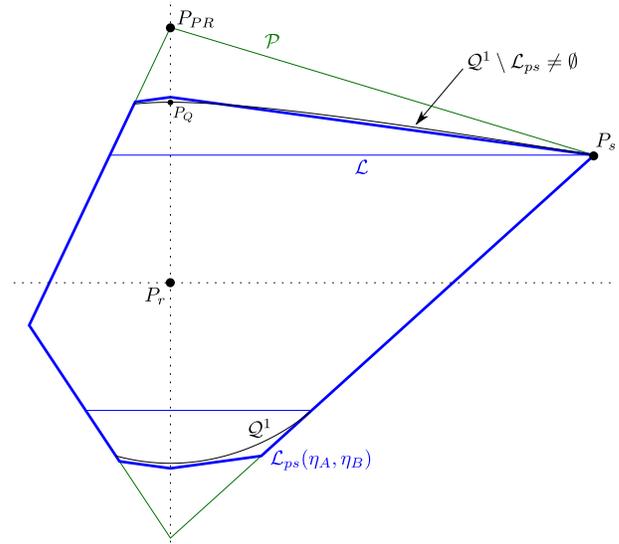}
\caption{(Color online.) Two-dimensional slice of the correlation space containing $P_{PR}$, $P_{s}$ and $P_r$. The thick blue lines delimit the post-selected local polytope ${\cal L}_{ps}(\eta_A,\eta_B)$; the thin blue lines delimit the local polytope $\cal L$, and the outer green lines delimit the non-signaling polytope $\cal P$. The quantum correlations are restricted by the black curves (which actually delimit the set ${\cal Q}^1 \supset {\cal Q}$ as defined in~\cite{navascues}). In this example, one can see that $P_Q$ is post-selected local, while there are still non-local correlations in ${\cal Q}^1 \setminus {\cal L}_{ps}$, closer to $P_s$ and to the boundaries of $\cal P$.
}
\label{fig_slice_PR_Ps}
\end{center}
\end{figure}

To illustrate this further, let us consider the 2-D slice containing $P_{PR}, P_r$ and
\ba
&
 P_{s} =
 \begin{array}{c||cc} 
 & 0 & 0 \\ \hline \hline
 0 & 0 & 0 \\
 0 & 0 & 0
 \end{array} \ ,
\ea
which is depicted in Figure~\ref{fig_slice_PR_Ps}. One clearly sees that for decreasing detection efficiencies, the correlation $P_Q$ becomes post-selected local before other correlations closer to $P_s$; the most robust correlations, for $\eta_A=\eta_B \to 2/3$, are those in the vicinity of $P_s$, in accordance with Eberhard's result~\cite{eberhard}.

\bigskip

It is interesting finally to compare the effect of post-selection with that of noise, another experimental imperfection. The effect of noise is simply to shrink the set of achievable correlations (for white noise, it simply corresponds to an homothetic transformation), without modifying the limits of the Bell local polytope $\cal L$; clearly, the most robust correlation to noise is the one that maximally  violates a Bell inequality. On the other hand, the effect of post-selection is to enlarge the set of (post-selected) local correlations, ${\cal L}_{ps}$. As we have seen, the transformation ${\cal L} \to {\cal L}_{ps}$ is not simply homothetic, and the previous intuition is not correct here.

\section{Conclusion}

We showed that the set of post-selected local correlations is a polytope, ${\cal L}_{ps}$, which can easily be derived from a larger polytope (the local {\it a priori} polytope ${\cal L}_0$), and which we could characterize in the CHSH scenario. In addition to providing Bell inequalities for post-selected correlations, our approach allowed us in particular to give a necessary condition on the detection efficiencies to be able to observe non-locality, and gave us a geometric intuition of the reason why the most robust correlation to detection inefficiencies is not the one that maximizes the violation of the standard CHSH inequality. 

Note that our approach directly gives a characterization of ${\cal L}_{ps}$ in terms of its facets; the difficulty, for larger numbers of inputs in particular, is to characterize the facets of ${\cal L}_0$. Another possibility would be to first characterize the vertices of ${\cal L}_{ps}$, and directly calculate its facets, without the need to evoke those of ${\cal L}_0$. It is unclear to us whether there is a way to do this more efficiently than with our approach.

Be that as it may, we believe that our approach should motivate the study of Bell polytopes for scenarios with more inputs, but where Alice and (/or~\footnote{The asymmetric case $\eta_A = 1$, $\eta_B < 1$ is indeed of particular interest for experiments using atom-photon entanglement~\cite{cabello_asym,brunner_asym}.}) Bob have 3 possible outputs, that would correspond to binary outcomes plus the no-detection possibility. Even if the local {\it a priori} polytope is not fully characterized, its known facets may imply non-trivial Bell-type inequalities for the corresponding post-selected local polytope.
In our study of the CHSH scenario, we found that all the facets of ${\cal L}_{ps}$ could be obtained from those delimiting ${\cal L}$, by simply grouping the no-detection events with another outcome. However, this does not hold in general, as we show in Appendix~\ref{app_3223}. It would be interesting to find other cases where ${\cal L}_{ps}$ has genuinely new facets compared to ${\cal L}$, and even find cases where these new facets can tolerate lower detection efficiencies to be violated.

Let us finally come back to the assumptions (\ref{eq_etaA}-\ref{eq_etaB}), that the detection efficiencies are independent of the choice of measurement settings. These assumptions were useful to carry out the present theoretical study, but might not be strictly satisfied in practical experiments. For practical purposes, one can either adapt our study to the observed situation, or simply avoid the detection loophole problem by not post-selecting the conclusive events, and consider the full {\it a priori} correlations directly.


\section{Acknowledgments}

I am grateful to Jean-Daniel Bancal, Nicolas Brunner, Eric Cavalcanti, Nicolas Gisin and Stefano Pironio for valuable discussions and comments. This work was supported by the Australian Research Council Centre of Excellence for Quantum Computer Technology.

\bigskip

\appendix

\section{Determining the facets of ${\cal L}_{ps}(\eta_A,\eta_B)$ in the CHSH scenario}

\label{app_facets_Lps}



\subsection{Sorting all the valid inequalities for ${\cal L}_{ps}$ \newline obtained from the facets of ${\cal L}_0$}

\label{app_all_ineqs}

From the 1116 facets of ${\cal L}_0$ (as defined in the CHSH scenario with imperfect detection efficiencies), and using eqs (\ref{eq_etaA}-\ref{eq_etaB}) and (\ref{eq_Pps}), we obtain a list of 1116 valid inequalities for ${\cal L}_{ps}$ (some of them appearing several times). Note that because of the particular role played by the no-detection outcome ``$\emptyset$", equivalent facets of ${\cal L}_0$ do not necessarily define equivalent inequalities for ${\cal L}_{ps}(\eta_A,\eta_B)$.

Most of these inequalities cannot be violated by any non-signaling correlations, and are simply implied by the non-negativity of the probabilities $P(a,b|x,y)$~\footnote{These inequalities can indeed be written in the form $\sum c_i P(a,b|x,y) \geq 0$, with only non-negative coefficients $c_i$.}. In addition to these trivial inequalities, we obtain 3 new inequalities (together with all their equivalent versions):
\ba
 & \begin{array}{c}
 -1 \leq I_{CH}^{\eta_A,\eta_B}
 \leq 0 \\
\mathrm{with}
 \quad
 I_{CH}^{\eta_A,\eta_B} \ = \
 \begin{array}{c||cc} 
 & -\eta_B & 0 \\ \hline \hline
 -\eta_A & \eta_A \eta_B & \eta_A \eta_B \\
 0 & \eta_A \eta_B & -\eta_A \eta_B
 \end{array} \, , 
 \end{array} \qquad 
 \label{eqApp_CH_eta} \\
 & \!\!\!\! \begin{array}{c||cc} 
 & -\eta_A \eta_B & 0 \\ \hline \hline
 -\eta_A \eta_B & \eta_A \eta_B & \eta_A \eta_B \\
 0 & \eta_A \eta_B & -\eta_A \eta_B
 \end{array}
 \ \leq \ \eta_A (1-\eta_B) + \eta_B (1-\eta_A) \, . \nonumber \\ \label{eqApp_CGLMP_eta}
\ea

The inequalities in (\ref{eqApp_CH_eta}) can be obtained from the CHSH inequalities, by grouping, for each observable, the outcome ``$\emptyset$" with one of the other outcomes; the inequality (\ref{eqApp_CGLMP_eta}) is obtained from the CGLMP inequality~(\ref{def_CGLMP}).

Interestingly, one can easily see that the inequality (\ref{eqApp_CGLMP_eta}) is actually implied by the upper bound in (\ref{eqApp_CH_eta}):
\ba
 && \!\!\!\!\!\!\!\!\! \begin{array}{c||cc} 
 & -\eta_A \eta_B & 0 \\ \hline \hline
 -\eta_A \eta_B & \eta_A \eta_B & \eta_A \eta_B \\
 0 & \eta_A \eta_B & -\eta_A \eta_B
 \end{array} \nonumber \\
 && \qquad = I_{CH}^{\eta_A,\eta_B}
 \ + \ \begin{array}{c||cc} 
 & \eta_B (1-\eta_A) & 0 \\ \hline \hline
 \eta_A (1-\eta_B) & 0 & 0 \\
 0 & 0 & 0
 \end{array} \nonumber \\ \nonumber \\
 && \qquad \leq \ 0 + \eta_A (1-\eta_B) + \eta_B (1-\eta_A) \, ,
\ea
so it is actually sufficient to only consider the inequalities (\ref{eqApp_CH_eta}).

\subsection{Facets of ${\cal L}_{ps}(\eta_A,\eta_B)$}

\label{app_facets}

The polytope ${\cal L}_{ps}$ is of dimension 8. To determine whether the remaining relevant inequalities are facets of ${\cal L}_{ps}(\eta_A,\eta_B)$, we can try, for each inequality, to extract 8 affinely independent correlations in ${\cal L}_{ps}$ that saturate it.

\subsubsection{Trivial facets}

The 12 deterministic correlations $P$ such that $P(00|11)=0$ all saturate the trivial bound $P(00|11) \geq 0$, and are clearly in ${\cal L}_{ps}$. Furthermore, one can easily extract 8 of them that are independent. The inequality $P(00|11) \geq 0$ is therefore a facet of ${\cal L}_{ps}$.

Equivalently, all the trivial inequalities $P(a,b|x,y) \geq 0$ are facets of ${\cal L}_{ps}$.

\subsubsection{Facets of the form $I_{CH}^{\eta_A,\eta_B} \leq 0$}

Using the decomposition
\ba
I_{CH}^{\eta_A,\eta_B} &\!=\!& - \big[ (\eta_A+\eta_B - 3\eta_A\eta_B) P(11|11) \nonumber \\
& & \ \ \ + \eta_A(1-\eta_B)  P(12|11) + \eta_B(1-\eta_A)  P(21|11) \nonumber \\
& & \ \ \ + \eta_A\eta_B \big( P(12|12)+P(21|21)+P(11|22) \big) \big] \, , \nonumber
\ea
we first note that if $\eta_A+\eta_B \geq 3\eta_A\eta_B$, then the inequality $I_{CH}^{\eta_A,\eta_B} \leq 0$ becomes trivial, since the coefficients in front of the probabilities in the above decomposition are then all non-negative.

\bigskip

Let us then assume that $\eta_A+\eta_B < 3\eta_A\eta_B$. Consider for instance the following correlations, written in the notation of (\ref{def_notation}), which all saturate the bound $I_{CH}^{\eta_A,\eta_B} = 0$:
\ba
 &
 \begin{array}{c||cc} 
 & 0 & 0 \\ \hline \hline
 0 & 0 & 0 \\
 0 & 0 & 0
 \end{array} \ , \quad
 \begin{array}{c||cc} 
 & 0 & 0 \\ \hline \hline
 0 & 0 & 0 \\
 1 & 0 & 0
 \end{array} \ , \quad
 \begin{array}{c||cc} 
 & 0 & 1 \\ \hline \hline
 0 & 0 & 0 \\
 0 & 0 & 0
 \end{array} \ , \nonumber \\
 &
 \begin{array}{c||cc} 
 & 1/2 & 1/2 \\ \hline \hline
 1/2 & x & 1/2 \\
 1/2 & 1/2 & 0
 \end{array} \ , \quad
 \begin{array}{c||cc} 
 & 1/2 & 1/2 \\ \hline \hline
 1/2 & 1/2 & x \\
 1/2 & 1/2 & 0
 \end{array} \ , \quad
 \begin{array}{c||cc} 
 & 1/2 & 1/2 \\ \hline \hline
 1/2 & 1/2 & 1/2 \\
 1/2 & x & 0
 \end{array} \ , \nonumber \\
 &
 \begin{array}{c||cc} 
 & 1/2 & 1/2 \\ \hline \hline
 1/2 & 1/2 & 1/2 \\
 1/2 & 1/2 & 1/2-x
 \end{array} \ , \quad
 \begin{array}{c||cc} 
 & 1/2 & 1/2 \\ \hline \hline
 y & y & y \\
 1/2 & 1/2 & 0
 \end{array} \ ,
\ea
where $x = \frac{\eta_A+\eta_B-2\eta_A\eta_B}{2\eta_A\eta_B}$ and $y = \frac{(1-\eta_A)\eta_B}{2\eta_A(2\eta_B-1)}$ (note that $0 \leq x,y < \demi$).

These 8 correlations are all in ${\cal L}_{ps}$ (they satisfy all the inequalities that delimit ${\cal L}_{ps}$), and are independent. This proves that when $\eta_A+\eta_B < 3\eta_A\eta_B$, the inequalities of the form $I_{CH}^{\eta_A,\eta_B} \leq 0$ are facets of ${\cal L}_{ps}(\eta_A,\eta_B)$.

\bigskip

\subsubsection{Facets of the form $I_{CH}^{\eta_A,\eta_B} \geq -1$}

Using the following decompositions:
\ba
&& \!\!\!\!\!\!\!\! I_{CH}^{\eta_A,\eta_B} + 1 \nonumber \\ \nonumber \\
&& \!\!\!\!\!\!\! = \eta_A\eta_B \big( P(22|11) + P(11|12) + P(12|22) \big) \nonumber \\
&& \!\!\!\! + (1-2\eta_A)\eta_B P(12|21) + (1-\eta_A)\eta_B P(22|21) \nonumber \\
&& \!\!\!\! + (1-\eta_A)(1-\eta_B) P_A(1|1) + (1-\eta_B) P_A(2|1) \label{decomp1} \\ \nonumber \\
&& \!\!\!\!\!\!\! = (1-\eta_A)(1-\eta_B) P(11|11) + (\eta_A+\eta_B-1) P(22|11) \nonumber \\
&& \!\!\!\! + (1-\eta_A) P(22|21) + (1-\eta_B) P(22|12) \nonumber \\
&& \!\!\!\! + (\eta_A\eta_B-\eta_A+\eta_B) \big(P(11|12) + P(12|22) \big)/2 \nonumber \\
&& \!\!\!\! + (\eta_A\eta_B+\eta_A-\eta_B) \big(P(11|21) + P(21|22) \big)/2  \nonumber \\
&& \!\!\!\! + (2-\eta_A-\eta_B-\eta_A\eta_B) \big(P(21|12) + P(12|21) \big)/2 , \qquad \quad \label{decomp2} 
\ea
we first note that if $\eta_A+\eta_B+\eta_A\eta_B \leq 2$, then the inequality $I_{CH}^{\eta_A,\eta_B} \geq -1$ becomes trivial:
\begin{itemize}
\item if $\eta_A \leq \demi$, then the coefficients in front of the probabilities in the decomposition (\ref{decomp1}) are all non-negative;
\item if $\eta_B \leq \demi$, then there exists a similar decomposition as (\ref{decomp1}) with non-negative coefficients;
\item if both $\eta_A,\eta_B \geq \demi$, and if $\eta_A+\eta_B+\eta_A\eta_B \leq 2$, then the coefficients in front of the probabilities in the decomposition (\ref{decomp2}) are all non-negative.
\end{itemize}

\bigskip

Let us refine the analysis in the case $\eta_A+\eta_B+\eta_A\eta_B > 2$, and define
\ba
f_1(\eta_A,\eta_B) &=& (1-\eta_A)(1-\eta_B)(3\eta_A\eta_B-\eta_A-\eta_B) \, , \nonumber \\
f_2(\eta_A,\eta_B) &=& f_1(\eta_A,\eta_B) + 2(1-\eta_A)^2\eta_B^2 \, , \nonumber \\
g(\eta_A,\eta_B) &=& f_1(\eta_A,\eta_B) + f_2(\eta_A,\eta_B) + f_2(\eta_B,\eta_A) \nonumber \\ && \quad + 2(1-\eta_A)(1-\eta_B)(\eta_A+\eta_B-2\eta_A\eta_B) \, , \nonumber \\
h(\eta_A,\eta_B) &=& f_1(\eta_A,\eta_B) + \frac{f_2(\eta_A,\eta_B)}{1-\eta_A} + \frac{f_2(\eta_B,\eta_A)}{1-\eta_B} \nonumber \\ && \quad - 2(\eta_A+\eta_B-2\eta_A\eta_B)(3\eta_A\eta_B-1) \, . \nonumber \\ \label{eq_def_fgh}
\ea
We then have
\ba
&& \!\!\!\!\!\!\! g(\eta_A,\eta_B)(I_{CH}^{\eta_A,\eta_B} + 1) \nonumber \\
&& = (1-\eta_A)(1-\eta_B)h(\eta_A,\eta_B) - f_1(\eta_A,\eta_B) I_{CH_1}^{\eta_A,\eta_B} \nonumber \\
&& \qquad - f_2(\eta_A,\eta_B) I_{CH_2}^{\eta_A,\eta_B} - f_2(\eta_B,\eta_A) I_{CH_2^\top}^{\eta_A,\eta_B} \nonumber \\
&& \qquad \quad + \, 2\,\eta_A\eta_B(1-\eta_A)(1-\eta_B) \nonumber \\
&& \!\!\!\!\!\!\! \times \big[ \eta_A(1-\eta_B) P(12|22) + \eta_B(1-\eta_A) P(21|22) \nonumber \\
&& \!\!\! + (\eta_A+\eta_B-2\eta_A\eta_B) \big( P(22|11) + P(22|12) + P(22|21) \big) \big] \ , \nonumber \\ \label{decomp4} 
\ea
where $I_{CH_1}^{\eta_A,\eta_B}, I_{CH_2}^{\eta_A,\eta_B}$ and $I_{CH_2^\top}^{\eta_A,\eta_B}$ are three equivalent versions of $I_{CH}^{\eta_A,\eta_B}$, defined as
\ba
I_{CH_1}^{\eta_A,\eta_B} \ &=& \
 \begin{array}{c||cc} 
 \eta_A\eta_B - \eta_A - \eta_B & \eta_A\eta_B & (1-\eta_A)\eta_B \\ \hline \hline
 \eta_A\eta_B & -\eta_A \eta_B & -\eta_A \eta_B \\
 \eta_A(1-\eta_B) & -\eta_A \eta_B & \eta_A \eta_B
 \end{array} \, , \nonumber
 \\
I_{CH_2}^{\eta_A,\eta_B} \ &=& \
 \begin{array}{c||cc} 
 -\eta_B & \eta_B & 0 \\ \hline \hline
 \eta_A\eta_B & -\eta_A \eta_B & -\eta_A \eta_B \\
 -\eta_A(1-\eta_B) & -\eta_A \eta_B & \eta_A \eta_B
 \end{array} \, , \nonumber
 \\
I_{CH_2^\top}^{\eta_A,\eta_B} \ &=& \
 \begin{array}{c||cc} 
 -\eta_A & \eta_A\eta_B & -(1-\eta_A)\eta_B \\ \hline \hline
 \eta_A & -\eta_A \eta_B & -\eta_A \eta_B \\
 0 & -\eta_A \eta_B & \eta_A \eta_B
 \end{array} \nonumber
\ea
(in this notation, the value in the top left corner is simply to be added to the combination of probabilities).

Note that under the assumption that $\eta_A+\eta_B+\eta_A\eta_B > 2$, we have $f_2(\eta_A,\eta_B)$, $f_2(\eta_B,\eta_A) \geq f_1(\eta_A,\eta_B) \geq 0$, and $g(\eta_A,\eta_B) > 0$ (except for the simple case $\eta_A=\eta_B=1$, for which $g(1,1) = 0$). Furthermore, the last three lines in (\ref{decomp4}) contain only non-negative coefficients. One can thus see that if, in addition, $h(\eta_A,\eta_B) \geq 0$ (or $\eta_A=1$, or $\eta_B=1$), then the inequality $I_{CH}^{\eta_A,\eta_B} \geq -1$ is simply implied by the (facet) inequalities $I_{CH_1}^{\eta_A,\eta_B}, I_{CH_2}^{\eta_A,\eta_B}, I_{CH_2^\top}^{\eta_A,\eta_B} \leq 0$.

\bigskip

Let us finally assume that $h(\eta_A,\eta_B) < 0$, and $\eta_A,\eta_B < 1$. Consider then the following correlation:
\ba
 \begin{array}{c||cc} 
 & \frac{1+z}{2} + \frac{\eta_A(1-\eta_B)}{(1-\eta_A)\eta_B}z & \frac{1+z}{2} \\ \hline \hline
 \frac{1+z}{2} + \frac{(1-\eta_A)\eta_B}{\eta_A(1-\eta_B)}z & \frac{3\eta_A\eta_B-\eta_A-\eta_B}{\eta_A\eta_B} \frac{1-z}{2} & (1 + \frac{(1-\eta_A)\eta_B}{\eta_A(1-\eta_B)})z \\
 \frac{1+z}{2} & (1 + \frac{\eta_A(1-\eta_B)}{(1-\eta_A)\eta_B})z & \frac{1+z}{2}
 \end{array} \nonumber
\ea
with $z=\frac{f_1(\eta_A,\eta_B)}{g(\eta_A,\eta_B)}$. For this correlation, we obtain $I_{CH}^{\eta_A,\eta_B} = -1 + (1-\eta_A)(1-\eta_B)\frac{h(\eta_A,\eta_B)}{g(\eta_A,\eta_B)} < -1$. However, one can check that this correlation satisfies all the trivial inequalities, and all those of the form $I_{CH}^{\eta_A,\eta_B} \leq 0$.

We thus see that when $h(\eta_A,\eta_B) < 0$ and $\eta_A,\eta_B < 1$, the inequality $I_{CH}^{\eta_A,\eta_B} \geq -1$ is no longer simply implied by the trivial and the $I_{CH}^{\eta_A,\eta_B} \leq 0$ facet inequalities. Since, from the previous analysis of Appendix~\ref{app_all_ineqs}, the facets of ${\cal L}_{ps}(\eta_A,\eta_B)$ can only be of the trivial form, of the form $I_{CH}^{\eta_A,\eta_B} \leq 0$, or of the form $I_{CH}^{\eta_A,\eta_B} \geq -1$, we conclude that the inequalities $I_{CH}^{\eta_A,\eta_B} \geq -1$ are therefore, in this case, also facets of ${\cal L}_{ps}(\eta_A,\eta_B)$.

\bigskip

\begin{figure}
\begin{center}
\epsfxsize=8cm
\epsfbox{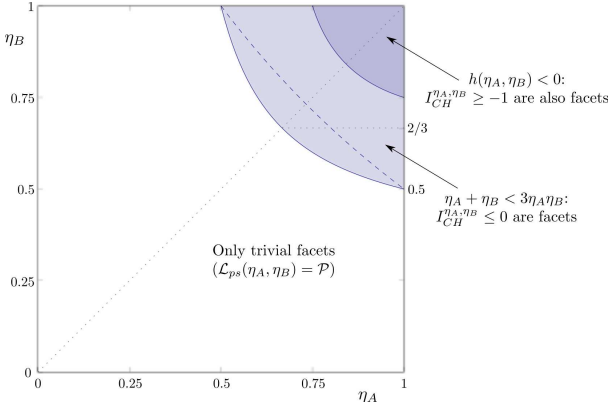}
\caption{Depending on the values of $\eta_A$ and $\eta_B$, different cases are encountered: in the white zone, only trivial inequalities delimit ${\cal L}_{ps}(\eta_A,\eta_B) = {\cal P}$; in the light gray zone, inequalities of the form $I_{CH}^{\eta_A,\eta_B} \leq 0$ are also facets, but not the inequalities $I_{CH}^{\eta_A,\eta_B} \geq -1$ (which can be violated by non-signaling correlations only above the dashed curve, i.e. for $\eta_A+\eta_B+\eta_A\eta_B > 2$); in the dark gray zone, both inequalities $I_{CH}^{\eta_A,\eta_B} \leq 0$ and $I_{CH}^{\eta_A,\eta_B} \geq -1$ are facets of ${\cal L}_{ps}(\eta_A,\eta_B)$.}
\label{fig_constr_etas}
\end{center}
\end{figure}

Figure~\ref{fig_constr_etas} illustrates the different cases that we studied, depending on the values of $\eta_A$ and $\eta_B$.

\section{${\cal L}_{ps}(1,\eta)$ for $m_A = 3$, $m_B=2$, $n_A=n_B=2$}

\label{app_3223}

Here we illustrate the fact that in general, the facets of ${\cal L}_{ps}$ can not all be derived from the facets of ${\cal L}$, by just grouping the no-detection events with another outcome.

To show that, it suffices to allow a third possible input for Alice ($m_A = 3, m_B=2$), all observable still having binary outcomes ($n_A=n_B=2$), and to consider the case when Alice has perfect detectors ($\eta_A=1$), while Bob's detection efficiency is $\eta < 1$.


\bigskip

We first note that in this scenario, the local polytope ${\cal L}$ only has trivial facets and facets of the CH form, where one of Alice's input is ignored~\cite{collins_gisin}.

The {\it a priori} polytope ${\cal L}_0$, corresponding to 3 and 2 inputs, 2 and 3 outputs for Alice and Bob respectively, can be characterized using standard polytope algorithms~\cite{lrs}. It is found to have 1260 facets, 36 of which are trivial, 216 are of the CH form (with one of Alice's inputs ignored, one of Bob's output group with another one), and respectively 288, 288 and 432 of them are of the following forms:
\ba
&
 I_{3223^{(1)}} \, =\ 
 \begin{array}{c||cc|cc} 
 & -1 & -1 & 0 & 0 \\ \hline \hline
 -1 & 1 & 1 & 1 & 0 \\ \hline
 0 & 1 & 0 & -1 & 0 \\ \hline
 0 & 0 & 1 & -1 & 0
 \end{array}
 \ \leq \ 0 \ , \label{def_I_3223_1}
\ea
\ba
 &
 I_{3223^{(2)}} \, =\ 
 \begin{array}{c||cc|cc} 
 & -1 & 0 & 0 & 0 \\ \hline \hline
 -1 & 1 & 0 & 1 & 0 \\ \hline
 -1 & 1 & 0 & 0 & 1 \\ \hline
 0 & 1 & 0 & -1 & -1
 \end{array}
 \ \leq \ 0 \ , \label{def_I_3223_2}
\ea
\ba
 &
 I_{3223^{(3)}} \, =\ 
 \begin{array}{c||cc|cc} 
 & 1 & -1 & 0 & 0 \\ \hline \hline
 1 & -1 & -1 & -1 & -1 \\ \hline
 0 & -1 & 1 & -1 & 1 \\ \hline
 0 & -1 & 1 & 1 & -1
 \end{array}
 \ \leq \ 1 \ . \label{def_I_3223_3}
\ea

Using similar methods as in Appendix~\ref{app_facets_Lps}, one can show that the facets of the corresponding post-selected local polytope ${\cal L}_{ps}(1,\eta)$ are either of the trivial form, or, if $\eta > \demi$, of either one of the two forms
\ba
I_{CH}^\eta \leq 0 \quad \mathrm{or} \quad I_{3223^{(1)}}^\eta \leq \eta \ ,
\ea
with
\ba
&
 I_{CH}^\eta \, =\ 
 \begin{array}{c||cc} 
 & -\eta & 0 \\ \hline \hline
 -1 & \eta & \eta \\
 0 & \eta & -\eta \\
 0 & 0 & 0
 \end{array} \ , \label{def_I_CH_eta} \\
&
 I_{3223^{(1)}}^\eta \, =\ 
 \begin{array}{c||cc} 
 & 0 & 0 \\ \hline \hline
 -(1-\eta) & 0 & \eta \\
 0 & \eta & -\eta \\
 \eta & -\eta & -\eta
 \end{array} \ . \label{def_I_3223_1_eta} \\ \nonumber
\ea

The inequalities $I_{CH}^\eta \leq 0$ can clearly be obtained from the CH inequalities that delimit ${\cal L}$, by grouping the no-detection events with another outcome. However, the inequalities $I_{3223^{(1)}}^\eta \leq \eta$ cannot be obtained from the facets of ${\cal L}$, and are genuinely new. This is in contrast with what we observed in the CHSH scenario, where the facets of ${\cal L}_{ps}$ could all be derived from the facets of ${\cal L}$.



Note finally that for all $\eta > \demi$, both types of inequalities can be violated by quantum correlations.
[To check that, one can consider for instance the following correlations: for $\eta > \demi$, define $X = \sqrt{\frac{\eta^2-(1-\eta)^2}{\eta^2+(1-\eta)^2}}$, $\theta = \arcsin X$ and $\ket{\psi} = \sin\frac{\theta}{2}\ket{00} + \cos\frac{\theta}{2}\ket{11}$; measuring $A_1=\sigma_z$, $A_2=\sigma_x$, $B_1=\frac{\sigma_z+X\sigma_x}{\sqrt{1+X^2}}$ and $B_2=\frac{\sigma_z-X\sigma_x}{\sqrt{1+X^2}}$ on $\ket{\psi}$ then gives $I_{CH}^\eta = \sqrt{\frac{\eta^2+(1-\eta)^2}{2}} - \demi > 0$, while measuring $A_1=\sigma_z$, $A_2=\frac{-\sigma_z+X\sigma_x}{\sqrt{1+X^2}}$, $A_3=\frac{-\sigma_z-X\sigma_x}{\sqrt{1+X^2}}$, $B_1=\sigma_x$ and $B_2=\sigma_z$ on $\ket{\psi}$ gives $I_{3223^{(1)}}^\eta = \eta + \sqrt{\frac{\eta^2+(1-\eta)^2}{2}} - \demi > \eta$.]

\end{document}